\documentclass[aps,prb,twocolumn,superscriptaddress]{revtex4}
\usepackage{graphicx}

\begin{document}

\title{Proximity induced metal/insulator transition in YBa$_{2}$Cu$_{3}$O$%
_{7}$/La$_{2/3}$Ca$_{1/3}$MnO$_{3}$ superlattices}
\author{Todd Holden}
\email{tholden@brooklyn.cuny.edu}
\affiliation{Max-Planck-Institut f\"{u}r Festk\"{o}rperforschung, Heisenbergstrasse 1,
D-70569 Stuttgart, Germany}
\affiliation{Physics Department, Brooklyn College of the City University of New York,
Brooklyn, New York 11210}
\author{H-U. Habermeier}
\author{G. Cristiani}
\affiliation{Max-Planck-Institut f\"{u}r Festk\"{o}rperforschung, Heisenbergstrasse 1,
D-70569 Stuttgart, Germany}
\author{A. Golnik}
\affiliation{Institute of Experimental Physics, Warsaw University, 69 Ho\.{z}a, 00-681
Warszawa, Poland}
\affiliation{Max-Planck-Institut f\"{u}r Festk\"{o}rperforschung, Heisenbergstrasse 1,
D-70569 Stuttgart, Germany}
\author{A. Boris}
\author{A. Pimenov}
\affiliation{Max-Planck-Institut f\"{u}r Festk\"{o}rperforschung, Heisenbergstrasse 1,
D-70569 Stuttgart, Germany}
\author{J. Huml\'{\i}\u{c}ek}
\affiliation{Department of Solid State Physics, Masaryk
University, Kotl\'{a}\v{r}sk\'{a} 2, CZ-61137 Brno, Czech
Republic} \affiliation{Max-Planck-Institut f\"{u}r
Festk\"{o}rperforschung, Heisenbergstrasse 1, D-70569 Stuttgart,
Germany}
\author{O. Lebedev}
\author{G. Van Tendeloo}
\affiliation{University of Antwerpen, EMAT RUCA, B-2020 Antwerp, Belgium}
\author{B. Keimer}
\author{C. Bernhard}
\affiliation{Max-Planck-Institut f\"{u}r Festk\"{o}rperforschung, Heisenbergstrasse 1,
D-70569 Stuttgart, Germany}
\date{\today }

\begin{abstract}
The far-infrared dielectric response of superlattices (SL) composed of
superconducting YBa$_{2}$Cu$_{3}$O$_{7}$ (YBCO) and ferromagnetic La$_{0.67}$%
Ca$_{0.33}$MnO$_{3}$ (LCMO) has been investigated by ellipsometry.
A drastic decrease of the free carrier response is observed which
involves an unusually large length scale of d$^{crit}\approx $20
nm in YBCO and d$^{crit}\approx $10 nm in LCMO. A corresponding
suppression of metallicity is not observed in SLs where LCMO is
replaced by the paramagnetic metal LaNiO$_{3}$. Our data suggest
that either a long range charge transfer from the YBCO to the LCMO
layers or alternatively a strong coupling of the charge carriers
to the different and competitive kind of magnetic correlations in
the LCMO and YBCO layers are at the heart of the observed
metal/insulator transition. The low free carrier response observed
in the far-infrared dielectric response of the magnetic
superconductor RuSr$_{2}$GdCu$_{2}$O$_{8}$ is possibly related to
this effect.
\end{abstract}

\maketitle

\section{Introduction}
The coexistence of such antagonistic phenomena as
superconductivity (SC) and ferromagnetism (FM) is a long-standing
problem in solid state physics. Originally it was believed that
they were mutually exclusive, but more recently it was found that
they can coexist under certain circumstances giving rise to novel
kinds of combined ground states \cite{Bulaevski}. Renewed interest
in SC and FM systems has been spurred by the search for novel
materials for applications in spintronic devices \cite{Wolf} as
well as by the observation that for a number of materials
(including the cuprate high-T$_{c}$'s) superconductivity occurs in
the vicinity of a magnetic instability \cite{Saxena1, Saxena2}.

Artificially grown heterostructures and superlattices (SLs) of
alternating SC and FM materials have become an important tool for
exploring the interplay between SC and FM. Of particular interest
have been SLs of perovskite-like transition metal oxides which
allow one to combine for example the cuprate
high T$_{c}$ superconductor (HTSC) YBa$_{2}$Cu$_{3}$O$_{7}$ (YBCO) with T$%
_{c}$=92 K with the manganite compound La$_{2/3}$Ca$_{1/3}$MnO$_{3}$ (LCMO)
that exhibits colossal magnetoresistance (CMR) and a FM ground state below T$%
^{Curie}$=240 K. The similar lattice constants and growth
conditions of YBCO and LCMO have enabled several groups to grow
SL's using various techniques like molecular beam epitaxy
\cite{Bozovic}, laser ablation \cite{HUH, Fabrega}, or magnetron-
and ion beam sputtering \cite{Yang, Jakob, Prieto, Przyslupski}.
Transport and magnetization measurements on these SL's have
established that there is a strong interaction between the SC and
FM order parameters in these SLs since both $T_{c}$ and $T_{mag}$
are considerably suppressed \cite{HUH,Prieto}. This suppression is
most pronounced for SLs with similarly wide YBCO and LCMO layers.
Notably, a sizeable suppression of $T_{c}$ and $T_{mag}$ was
observed even for SLs with relatively thick layers of
d$_{\text{YBCO}}$,d$_{\text{LCMO}}>$10nm. This observation implies
that the proximity coupling involves an unexpectedly large length
scale far in excess of the SC coherence length of $\xi _{SC}\leq
$2 nm. Equally remarkable are some reports of a considerable
suppression of the normal state electronic conductivity
\cite{Yang, Prieto, Przyslupski} which at a first glance is not
expected since these SLs are composed of metallic materials.

These puzzling observations motivated us to investigate the
electronic properties of SC/FM SLs by means of spectral
ellipsometry. Unlike transport measurements, this optical
technique is not plagued by contact problems and allows one to
reliably obtain the bulk electronic properties of a given material
since grain boundaries with lower conductivity or filamentary
pathes of least resistance do not contribute significantly. We
investigated the far-infrared dielectric properties of a series of
SLs that are composed of thin layers of YBCO and LCMO. Our optical
data provide clear evidence that the free carrier response in
these SC/FM SLs is strongly suppressed as compared to the pure
films of which they consist. The suppression appears in the normal
state as well as in the SC state. It depends strongly on the
thickness ratio, d$_{\text{YBCO}}$/d$_{\text{LCMO}},$ and is most
pronounced for a 1:1 ratio. Our most important observation is that
the length scale involved is surprisingly large with nearly
complete suppression for layer thicknesses of
d$_{\text{YBCO}}^{crit}\approx $20 nm and
d$_{\text{LCMO}}^{crit}\approx $10nm. A similar suppression is
observed for SLs that are composed of YBCO and the FM metal
SrRuO$_{3}$ (SRO). In stark contrast, we observe no corresponding
suppression of metallicity in similar SLs that consist either of
YBCO and the paramagnetic metal LaNiO$_{3}$ (LNO) or of the
insulating compound PrBa$_{2}$Cu$_{3}$O$_{7}$ (PBCO).

\begin{figure}[tbp]
\includegraphics[width=8.6cm]{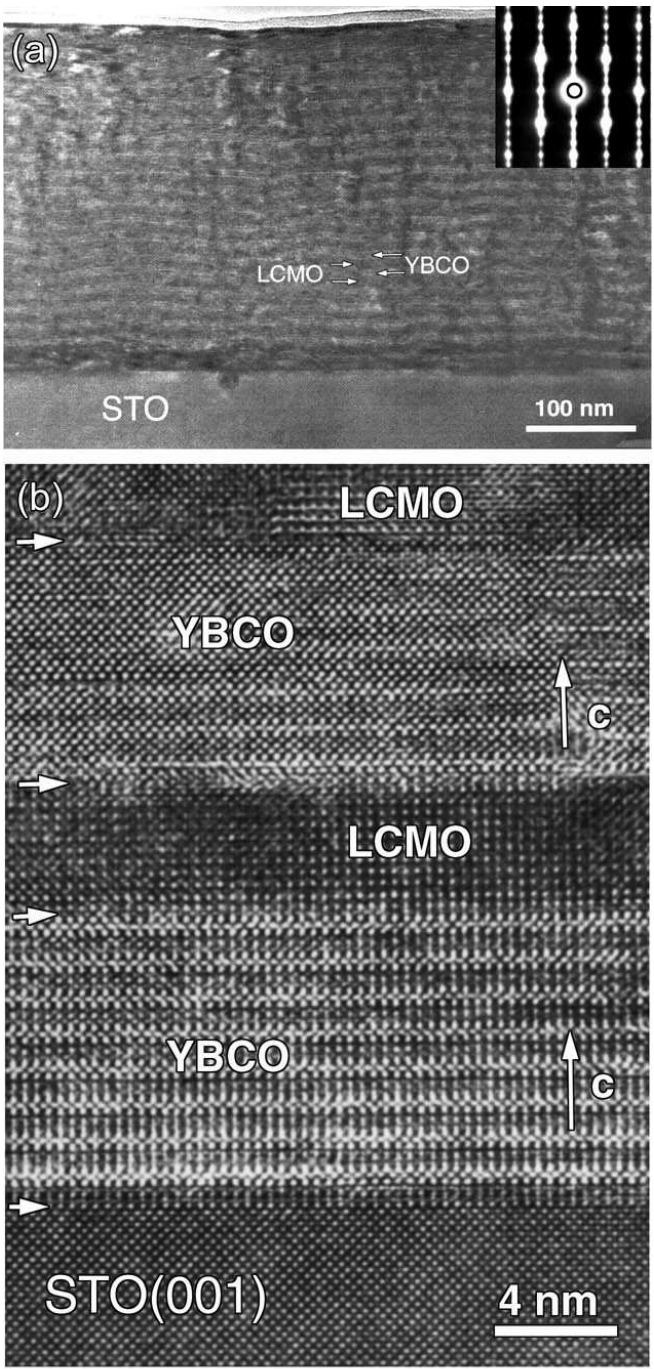}
\caption{(a) Low resolution (b) high resolution transmission
electron microscope and electron diffraction (shown on inset)
images of a [8nm:6nm]x20 superlattice. Note that the first YBCO
and LCMO layer thicknesses differ from the others.}
\end{figure}

\section{Experimental Details}

We have grown SLs of YBCO/LCMO, YBCO/SRO, YBCO/LNO, YBCO/PBCO and
also films of the pure materials by laser ablation on SrTiO$_{3}$
substrates as described in Ref. \onlinecite{HUH}. The composition
of the films and their high quality has been confirmed by x-ray
diffraction analysis, transmission electron microscopy (TEM) and
also by Raman measurements. A TEM image of a [8:6nm]x20 YBCO/LCMO
SL is displayed in Fig. 1. It shows that the interfaces are
atomically sharp and epitaxial. When viewed at low magnification
the interfaces appear somewhat wavy. A similar waviness is
commonly observed in SLs containing YBCO and most likely is
related to strain relaxation \cite{Vigliante}. The X-ray
diffraction patterns exhibit only the corresponding (00h) peaks
for YBCO, LCMO, and for the SrTiO$_{3}$ substrate confirming the
phase purity and the epitaxial growth of the SLs. The SL peaks are
not well resolved from the main peaks due to the low resolution of
the instrument and the interface waviness. The SC and the FM
transition temperatures as determined by measurements of the
dc-conductivity and SQUID magnetization are summarized in Table I.

The ellipsometric measurements have been performed with a
home-built setup at the U4IR and U10A beamlines of the National
Synchrotron Light Source (NSLS) in Brookhaven, USA and, in parts,
using the conventional mercury arc lamp of a Bruker 113V FTIR
spectrometry \cite{Henn, Golnik}. In practice, the
pseudo-dielectric function for an anisotropic crystal deviates
only slightly from
the actual dielectric function along the plane of incidence (ab-plane in our case) %
\cite{Kircher, Henn}. The spectra were analyzed with a multilayer
ellipsometric analysis program. The film thickness was refined by
minimizing features in the calculated film pseudo-dielectric
function that arise from the phonons of the SrTiO$_{3}$ substrate.
It was generally found to agree well with the nominal thickness
based on the growth conditions and the TEM data.
In many spectra, small artifacts remain due to the Berreman mode near 480 cm$%
^{-1}$ and the STO phonons near 170 and 550 cm$^{-1}$, due to
small differences of our substrates from the STO reference
\cite{Henn, Kamaras}. Since the SL thickness is well below the FIR
wavelength, the entire SL can be treated as a single layer
according to effective medium theory. Accordingly an effective
dielectric function can be obtained which, for this geometry,
corresponds to the volume average of the dielectric functions of
the components of the SL \cite{Aspnes}. Below we will use
$\varepsilon _{1}$ and $\sigma _{1}$ to denote the a SL's
effective dielectric function and corresponding effective
conductivity when discussing SLs.

\begin{table*}[tbp]
\caption{Physical parameters for representative SLs and Films grown by Laser
Ablation}
\label{Superlattice parameters}
\begin{ruledtabular}
\begin{tabular}{lcccccccc}
\textbf{\lbrack d}$_{\textbf{YBCO}}$\textbf{:d}$_{\textbf{LCMO}}$\textbf{%
]} & $T_{c}$ & $T_{mag}$ &
      \textbf{$\omega_{p}^{2}$(10K)}& \textbf{$\Gamma$(10K)}& \textbf{$\omega_{p}^{2}$(100K)} & \textbf{$\Gamma$(100K)}& \textbf{$\omega_{p}^{2}$(300K)}&\textbf{$\Gamma$(300K)}\\
\ &\textbf{(K)} &\textbf{(K)} & \textbf{(eV}$^{2}$\textbf{)} & \textbf{(eV}$^{2}$\textbf{)}& \textbf{(meV)} & \textbf{(meV)} \\
\lbrack 8nm:6nm]x20 & 60 & 120 & 0.035 &49&0.029&49& 0.024 &50\\
\lbrack 5nm:5nm]x40 & 60 & 120 & 0.026 & 31 & 0.025 &28&0.025&21\\
\lbrack 16nm:16nm]x10 & 73 & 215 & 0.36&26&0.29&33 & 0.11 &32\\
\lbrack 60nm:60nm]x5 & 85 & 245 & 0.63 &22&0.55&33& 0.37 &66\\
\lbrack 60nm:15nm]x5 & 86 & 160 & 1.44&27&1.41&49 & 1.14 &79\\
\lbrack 30nm:15nm]x5 & 80 & 165 & 0.80 & 27 & 0.84 & 43 & 0.7 &65\\
\lbrack 13nm:5nm]x20 & 56 & 115 & 0.44 & 29 & 0.43 & 38 & 0.36 &66\\
\lbrack 8nm:3nm]x20 & 60 & 120 & 0.55 & 34 & 0.55 & 43 & 0.46 &54\\
\lbrack 15nm:30nm]x5 & - & 195 & 0.39 & 44 & 0.22 & 44 & 0.12 & 38\\
\lbrack 15nm:60nm]x5 & - & 240 & 1.03 & 42 &0.80 & 43 & 0.064 &21\\
\textbf{\lbrack d}$_{\textbf{YBCO}}$\textbf{:d}$_{\textbf{LNO}}$\textbf{]%
} &  &  &  \\
\lbrack 5nm:5nm]x20 & 33 & - & 1.15 & 69 & 1.10 & 72 & 1.04& 76\\
\lbrack 10nm:10nm]x20 & 70 & - & 1.19 & 57 & 1.21& 60&1.07 & 72\\
\textbf{\lbrack d}$_{\textbf{YBCO}}$\textbf{:d}$_{\textbf{PBCO}}$\textbf{]%
} &  &  &  \\
\lbrack 10nm:10nm]x20 & 85 & - & 0.36 & 12 & 0.57 & 31 & 0.49 & 48\\
\textbf{Pure Materials} &  &  &  \\
YBCO & 90 & - & 0.93 & 19 & 1.22 & 42 & 1.03 &75\\
LCMO & - & 245 & 1.08 & 37 & 0.61 & 37 & 0.03 &10\\
LNO & - & - & 0.99 & 103 & 0.95 & 98 & 1.04 & 111\\
Ru-1212 & - & 145 & 0.30 & 28 & 0.28 & 32 & 0.24 & 53\\
\end{tabular}
\end{ruledtabular}
\end{table*}

\begin{figure}[tbp]
\caption{in-plane conductivity, $\protect\sigma _{1}$, and the
dielectric function, $\protect\varepsilon _{1}$ for representative
SLs with double layers of (a) 60nm:60nm, (b) 16nm:16nm, (c)
8nm:6nm, and (d) 5nm:5nm. (d) Numerical simulation for a SL with
bilayers of 16nm normal LCMO and 16 nm fit layer with
$\protect\omega_{p}^{2}$=0.03 eV$^{2}$ [similar to (c) and (d)]}
\label{}
\includegraphics[width=8.6cm]{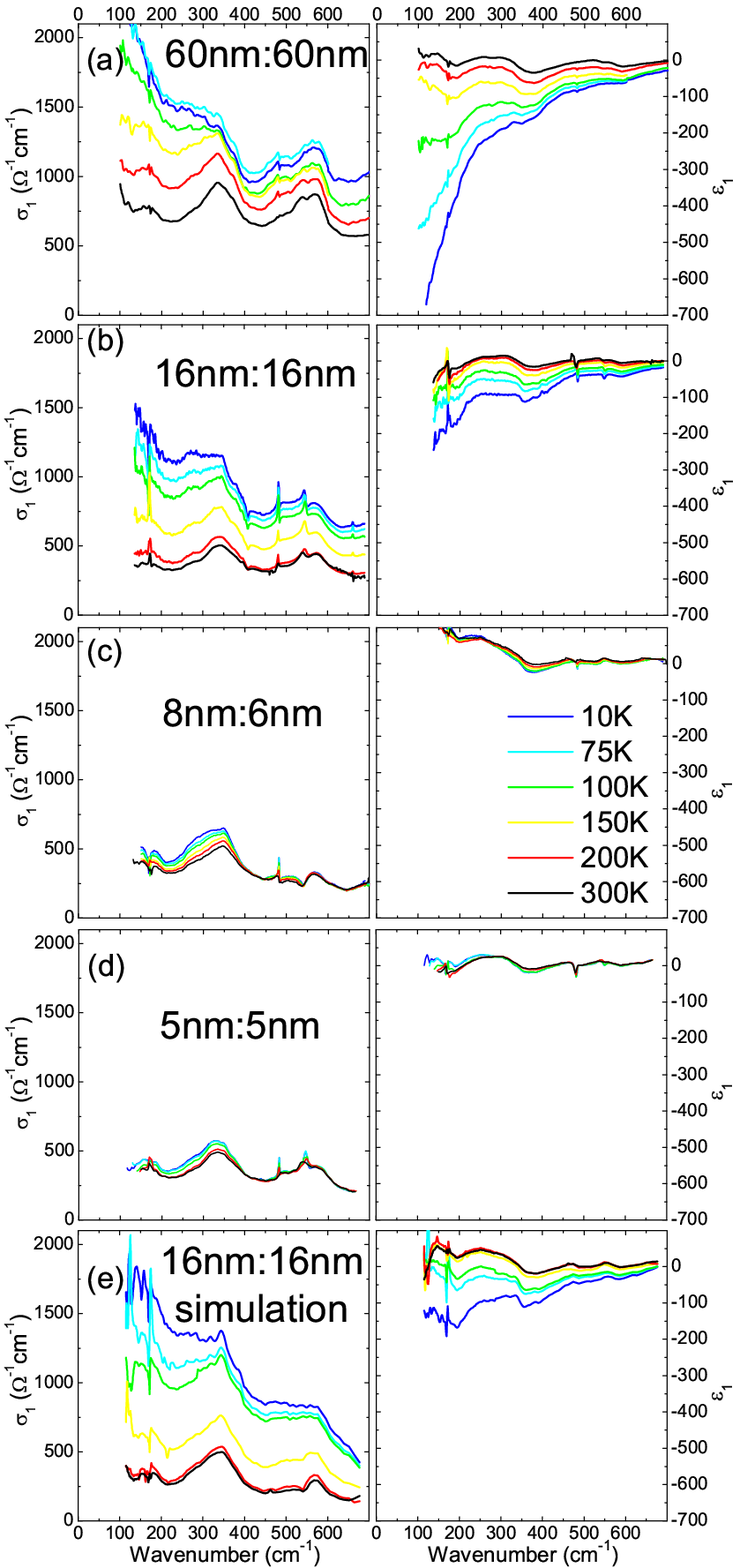}
\end{figure}

\section{Results and Discussion}

Figure 2 shows representative spectra for the real-parts of the in-plane
conductivity, $\sigma _{1}$, and the dielectric function, $\varepsilon _{1}$%
, of several YBCO/LCMO SLs with a thickness ratio close to 1:1, of
(a) 60:60 nm, (b) 16:16 nm, (c) 8:6 nm, and (d) 5:5 nm. Shown are
spectra in the normal and in the SC state. Given the metallic
properties of the pure YBCO and LCMO films (spectra are not shown)
one would expect that the SLs also should exhibit a strong
metallic response. Instead Figure 2(a-d) highlights that the
YBCO/LCMO SLs exhibit a drastic decrease in the absolute value of
$\sigma _{1}$ and $\varepsilon _{1}$ which corresponds to a
significant reduction of the free carrier concentration or of
their mobility. This suppression of metallicity is still fairly
weak for the 60:60nm SL but becomes sizeable already for the
16:16nm SL. Finally, for the 8:6nm and 5:5nm SLs, the free carrier
response is barely visible and the spectra are dominated by phonon
modes that are characteristic for LCMO and YBCO. We only note here
that we observe a similar effect for the SLs of YBCO/SRO which
contain the FM metal SRO.

For a quantitative description of the free carrier response we
have modelled the spectra with a Drude-function plus a sum of
Lorentzian functions that account for the phonon modes and the
so-called MIR-band at higher frequencies. The square of the
extracted plasma frequency, $\omega _{p}^{2}=4\pi n/m$*, which is
proportional to the ratio of the free carrier concentration, n,
divided by their effective mass, $m$*, are given in table I (at
10K, 100K, and 300 K). The value of $\omega _{p}^{2}$ is
proportional to the free carrier spectral weight (SW), which is
the dominant contribution to the area under the $\sigma _{1}$
curve in the FIR. Also shown is the scattering rate, $\Gamma $,
which accounts for the broadening of the Drude-response due to
scattering of the charge carriers. The results are representative
for a significantly larger number of SLs that have been
investigated. The value of $\omega _{p}^{2}$ can be seen to
decrease by more than an order of magnitude as the layer thickness
is reduced from 60:60nm to 8:6nm. However, even the 8:6 nm SL,
despite its very low $\omega _{p}^{2}$ and the correspondingly low
density of the SC condensate, exhibits a superconducting
transition in the measured resistivity at $T_{c}$=60 K. At
the same time this SL still exhibits a ferromagnetic transition at $T_{mag}$%
=120 K. A significant suppression of $\omega _{p}^{2}$ is evident
already for the 16:16nm SL. This effect is most pronounced at 300
K, i.e. above the CMR transition at $T_{mag}$=215 K where the LCMO
layers are known to remain insulating. The apparent increase in
conductivity below 200 K is coincident with the FM transition and
thus with the well known MIT transition in the LCMO layers that is
at the heart of the CMR effect. This finding suggests that the
metallicity of the YBCO layers is already almost entirely
suppressed for the 16:16nm SL whereas the LCMO layers still become
metallic below the FM transition. To confirm this interpretation,
we fitted the response of the 16:16nm SL using the response
functions of pure YBCO and LCMO layers (as measured by
ellipsometry), as well as a theoretical fit function with a Drude
plus a broad Lorentzian term to account for the so-called
mid-infrared band.\ As shown in Fig. 2(e) we obtained a good fit
at all temperatures for a model where the SL is composed of 16 nm
LCMO and 16 nm of a fit layer with $\omega _{p}^{2}$ = 0.03
eV$^{2}$ (as in the 8nm:6nm SL). We were not able to fit the data
with a model SL of 16 nm YBCO and 16nm fit layer.

A corresponding suppression of metallicity is not observed for a
SL where the FM metallic LCMO layers are replaced by layers of insulating PrBa$_{2}$Cu$%
_{3}$O$_{7}$ or LaNiO$_{3}$ (LNO), a paramagnetic metal (PM) that
is characterized by a broad Drude-peak and a strong electronic
mode around 300 cm$^{-1}$\cite{Massa}. Figure 3(a) displays our
ellipsometric data on a 5:5 nm SL of YBCO/LNO. It is immediately
evident that this sample (despite of its very thin individual
layers) maintains a metallic response with $\omega _{p}^{2}$ =
1-1.2 eV$^{2}$. The apparent broadening of the Drude-response of
this SL with $\Gamma\approx$ 70 meV is partly due to the broad
nature of the Drude response in LNO, but may also be caused by the
waviness of the very thin layers or possibly also by the diffusion
of a minor amount of Ni from the LNO to the YBCO layer. This
effect may also be responsible for the sizeable suppression of
T$_{c}$. Figure 3(b) shows that a similar
persistence of metallicity is evident for a SL with 10:10 nm of YBa$_{2}$Cu$%
_{3}$O$_{7}$/PrBa$_{2}$Cu$_{3}$O$_{7}$ (YBCO/PBCO). Since it is well known
that the PBCO layers are in an insulating state, it it is clear that the
free carrier response arises solely due to the metallic YBCO layers here.

\begin{figure}[tbp]
\caption{in-plane conductivity, $\protect\sigma _{1}$, and the
dielectric function, $\protect\varepsilon _{1}$ for representative
SLs with double layers of (a) [5nm YBCO:5nm LNO]x20 and (b) [10nm
YBCO:10nm PBCO]x20.}
\label{}
\includegraphics[width=8.6cm]{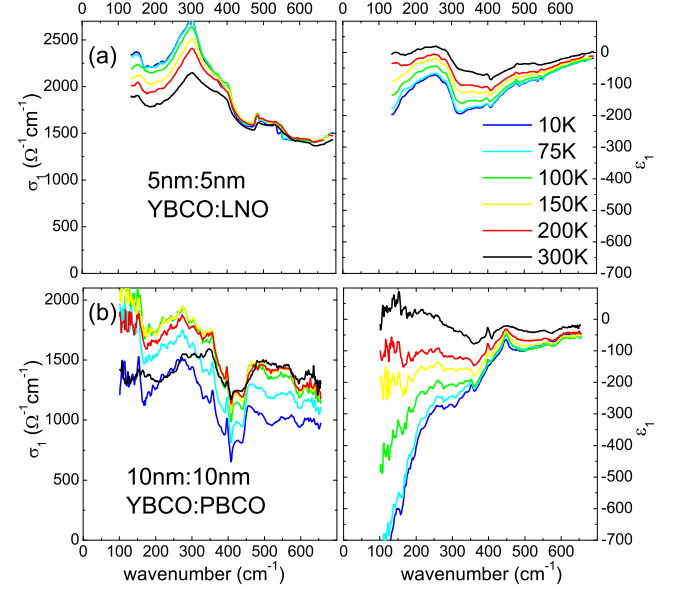}
\end{figure}

Even for the YBCO/LCMO SLs we find that the metallic response can
be recovered by changing the thickness ratio in favor of the YBCO\
layers. Figure 4 shows optical spectra on representative YBCO/LCMO
SLs with a thickness ratio close to 3:1, for (a) 60:15nm, (b)
30:15nm, (c) 13:5nm and (d) 8:3nm. It is immediately evident that
the 3:1 SLs exhibit a much weaker suppression of $\omega _{p}^{2}$
than the 1:1 SLs shown in Fig. 2. Most instructive is the large
difference between the 8:3nm and the 8:6nm SLs in Fig. 2(c) and
3(b). While the 8:6nm SL exhibits nearly insulating behavior, the
signature of a sizeable free carriers response is clearly evident
for the 8:3 nm SL. Such a result excludes any kind of structural
or chemical imperfections of the SLs, such as the roughness of the
interfaces or some kind of diffusion of the cations of the LCMO
layer across the interfaces as a possible origin for the
suppression of metallicity in the YBCO layer. As mentioned above,
a poor material quality or a significant chemical mixing across
the layer boundaries is furthermore excluded by our x-ray, TEM and
also by preliminary secondary ion mass spectrometry (SIMS)
experiments. Furthermore, these problems should be even more
severe for the 5:5nm YBCO/LNO SL which remains metallic.

A possible explanation of the dramatic suppression of metallicity in the 1:1
SC/FM SLs would be a massive transfer of holes from the YBCO layers to the
LCMO layers. Such a transfer of about 3x10$^{21}$ holes/cm$^{3}$ could
severely deplete the YBCO layers and, according to the phase diagram of LCMO %
\cite{Schiffer}, could drive the LCMO layers into a charge ordered
state similar to the one observed for a Ca content of x$>$0.45. If
this is the case, the LCMO is acting somewhat like an n-type
semiconductor by accepting holes; however, the implied phase
change to a charge ordered state differentiates this from a
classical p-n junction, in addition to the large charge density
involved. At a first glance one might think that such a scenario
is not very likely. According to the simple depletion layer model
of semiconductor theory\cite{Ashcroft}, the Poisson equation is
solved approximately with a quadratic potential difference on both
sides of the interface giving a total depletion layer $ d =
\sqrt{2\varepsilon \Delta \phi / (\pi Ne)} $. Here $N$ is the
volume density of free carriers (assumed to be equal on the two
sides of the interface), $\Delta\phi$ is the potential difference
between the bulk and interface, $\varepsilon$ is the static
dielectric constant of the depleted insulating material
(consisting of the electronic and phononic contributions), and $e$
is the electron charge. Given a difference in work functions of 1
eV, and assuming a fairly large value $\varepsilon$=15, one
expects the depletion layer of about 1 nm thickness, i.e., 1
monolayer of YBCO. However, estimates using the Lindhard
dielectric function\cite{Ashcroft} for an anisotropic degenerate
fermi--gas, parameterized so as to be representative of the
layered YBCO, predict thicker depleted regions with Friedel
oscillations of the charge density along the $c$--axis. The charge
redistribution might actually affect several monolayers of
(otherwise optimally doped) copper--oxygen layers in YBCO. Note
that the charge depletion in infinite SL's would be symmetric at
both interfaces of the YBCO layers.

\begin{figure}[tbp]
\includegraphics[width=8.6cm]{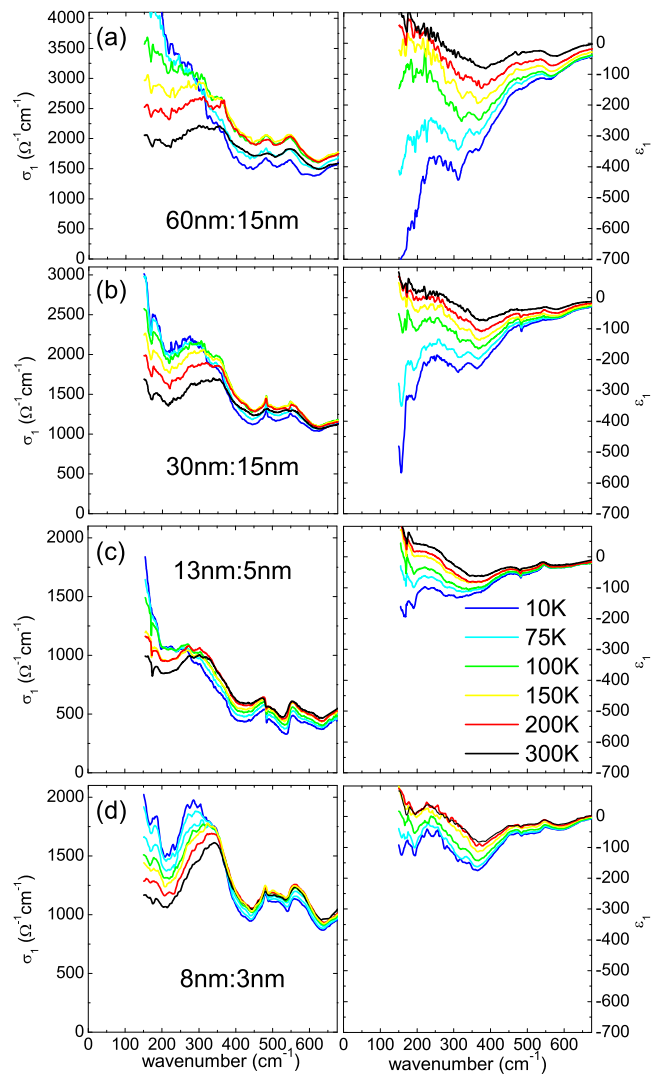}
\caption{in-plane conductivity, $\protect\sigma _{1}$, and the
dielectric function, $\protect\varepsilon _{1}$ for representative
SLs with YBCO/LCMO ratios close to 3:1 (a) 60nm:15nm, (b)
30nm:15nm, (c) 13nm:5nm, and (d) 8nm:3nm.}
\end{figure}

Another equally interesting possibility is motivated by our
observation that a suppression of metallicity occurs only in case
of the FM layers LCMO and SRO whereas it is absent for the
paramagnetic metal LaNiO$_{3}$. This suggests that magnetic
correlations play an important role in the observed
metal/inuslator transition, possibly due to a novel magnetic
proximity effect where charge carriers that are strongly coupled
to different and competitive kinds of magnetic correlations, i.e.
FM ones in the LCMO as opposed to AF or more exotic ones in YBCO,
become localized. The underlying idea would be that the charge
carriers gain mobility by adjusting their spins to the
corresponding magnetic background, i.e. to the Cu moments in YBCO
and the Mn(t$_{2g}$) moments in LCMO. Such a scenario is already
well established for the case of LCMO where it leads to the well
known CMR effect. For YBCO, however, this is not the case.
Nevertheless, it is known that AF correlations and fluctuations
persist even for optimally doped samples. There exists clear
evidence that the charge dynamics is strongly affected by the
magnetic correlations, the most prominent example is the so-called
pseudogap phenomenon in underdoped samples. Indeed, a number of
models have been proposed where the mobility of the charge
carriers strongly depends on the magnetic correlations and where a
transition to a nearby insulating ground state can be induced by
magnetic interactions, including the stripe phase \cite{Demler,
Emery97}, RVB-type \cite{Anderson}, SO(5) \cite{Zhang}, and the
phase separation \cite{Burgy} models. The effect of a proximity
coupling to a metallic FM layer has not been considered yet for
any of these models.

In this context the most important aspect concerns the
unexpectedly large length scale that is involved in the
suppression of conductivity. There is indeed experimental
indication that the spin coherence length in the cuprate HTSC is
unusually large of the order of 20 nm \cite{Wei} or more
\cite{Pai}. In addition, LCMO has a finite DOS near the Fermi
level for both spin polarizations, although the spin mobility is
much higher for the majority spins \cite{Nadgorny}. Therefore it
is not impossible that spin diffusion (driven by the gradient in
spin polarization between LCMO and YBCO and opposed by the
relaxation in the YBCO layer) may lead to a long-range spin
polarization of the charge carriers deep inside the YBCO layers.
Alternatively, the yet unknown novel magnetic ground state of the
underdoped and optimal doped cuprate HTSC may be associated with
an unusually large coherence length. Evidence for a long-range
proximity effect has indeed been recently obtained in photo-doped
YBa$_{2}$Cu$_{3}$O$_{6}$ \cite{Decca}, where Josephson-tunneling
currents were observed across undoped (AF) regions as wide as 100
nm.

\begin{figure}[tbp]
\includegraphics[width=8.6cm]{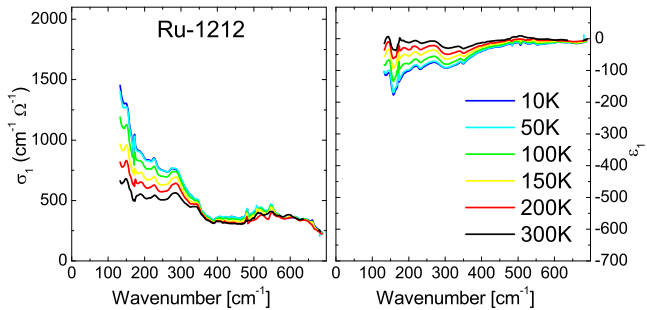}
\caption{in-plane conductivity, $\protect\sigma _{1}$, and the
dielectric function, $\protect\varepsilon _{1}$ for a laser
ablation grown RuSr$_{2}$GdCu$_{2}$O$_{8}$ film.}
\end{figure}

Clearly, further experiments are required \ before one can distinguish
between these equally fascinating possibilities. Most important will be
direct measurements of the hole content within the CuO$_{2}$ planes which
can be performed for example with the technique of core-level spectroscopy.
Further attempts should include studies of the field-effect or of
photo-induced conductivity as well as optical measurements in applied
magnetic fields.

Finally, we make a comment on the infra-red conductivity of the hybrid
ruthenate-cuprate compound RuSr$_{2}$GdCu$_{2}$O$_{8}$ (Ru-1212), in which
SC within the CuO$_{2}$ layers ($T_{c}$=50 K) and strong magnetism (with a
sizeable FM component) in the RuO layers ($T_{mag}$=135 K) can coexist
within a unit cell \cite{Bernhard99}. Thus in some sense it is a
cuprate/magnetic SL, similar to these YBCO/LCMO SLs, with layer thicknesses
of only a few angstrom. It is still debated whether the interaction between
the SC and the magnetic order parameters is weak (this may be possible due
to the layered structure), or whether their coupling is strong and therefore
gives rise to a novel ground state with interesting new properties. Indeed
some experiments indicate that the same charge carriers, which eventually
become SC below $T_{c}$, are strongly coupled to the Ru magnetic moments %
\cite{McCron,Tokunaga}. Another unusual feature of Ru-1212 is that
it is a surprisingly poor conductor with a low dc conductivity and
extremely small SC condensate density as compared to other HTSCs
\cite{Bernhard00}. Figure 5 shows the infrared conductivity and
dielectric function of a laser ablation grown Ru-1212 thin film.
Raman and x-ray characterization of this film show it to be
$\approx 95\%$ phase pure with the c-axis along the growth
direction. Based on SQUID magnetization measurements the magnetic
ordering transition of the Ru-moments occurs at T$^{mag}$=145 K
and there is no evidence for superconductivity in this particular
film. In fact it is commonly found for Ru-1212 that bulk
superconductivity occurs only in samples with T$_{mag}\leq $135 K.
Evidently, the free carrier response of this film with $\omega
_{p}^{2}$ $\leq$ 0.3 eV$^{2}$  is much weaker than that of YBCO.
The analogy to our artificial YBCO/LCMO SLs is rather striking and
suggests that a related effect may be at work in the compound,
which can be viewed as an intrinsic SL of a superconducting and
magnetic layers.

\section{Summary and Conclusions}

In conclusion, we have reported ellipsometric measurements of the
far-infared (FIR) dielectric properties of super-lattices (SLs) composed of
thin layers of YBa$_{2}$Cu$_{3}$O$_{7}$ (YBCO) and La$_{0.67}$Ca$_{0.33}$MnO$%
_{3}$ (LCMO) that have been grown by laser ablation. Our optical data
provide clear evidence that the free carrier response is strongly suppressed
in these SLs as compared to the one in the pure YBCO and LCMO films. The
suppression occurs in the normal as well as in the SC state and it involves
a surprisingly large length scale of the order of d$_{\text{YBCO}}^{crit}$%
=20 nm and d$_{\text{LCMO}}^{crit}$=10nm. A similar suppression is
observed for YBCO/SrRuO$_{3}$ SLs. In stark contrast, a
corresponding suppression of free carrier response does not occur
for SLs where the FM LCMO is replaced by the paramagnetic metal
LaNiO$_{3}$. Possible explanations have been discussed in terms of
a charge transfer between adjacent layers as well as charge
localization due to magnetic correlations that are induced by a
novel kind of long-range proximity effect. The low free carrier
response observed
in the far-infrared dielectric response of the magnetic superconductor RuSr$%
_{2}$GdCu$_{2}$O$_{8}$ is possibly related to this effect.

\begin{acknowledgments}

T.H. gratefully acknowledges the support of the Alexander von Humboldt
Foundation. For technical help at the NSLS we thank L.G. Carr and C.C.
Homes. The technical support by R.K. Kremer, E. Br\"{u}cher, A. St\"{a}rke
at MPI-FKF is highly appreciated. Some ellipsometry measurments have been
performed by Julia Greisl from California Technical Institute during here
stay at MPI-FKF.

\end{acknowledgments}


\bigskip



\end{document}